\documentclass[letterpaper, 10 pt, conference]{ieeeconf}  

\IEEEoverridecommandlockouts                              

\usepackage{graphicx} 
\usepackage{amsmath} 
\usepackage{amssymb}  
\usepackage{xcolor}
\usepackage{bm}
\usepackage{placeins}
\usepackage{lipsum,graphicx,subcaption}
\usepackage{pgfplots, tikz}

\pgfplotsset{compat=1.3}

\title{\LARGE \bf
3D Localization of a Sound Source Using Mobile Microphone Arrays Referenced by SLAM
}

\author{Simon Michaud, Samuel Faucher,  Fran\c{c}ois Grondin, Jean-Samuel Lauzon, Mathieu Labb\'{e}, \\ Dominic L\'{e}tourneau, Fran\c{c}ois Ferland, Fran\c{c}ois Michaud
\thanks{This work was supported in part by the Natural Sciences and Engineering Research Council of Canada (NSERC), the Fonds de recherche du Qu\'{e}bec - Nature et technologies (FRQNT) and ACELP-3IT Funds, Universit\'{e} de Sherbrooke.}
\thanks{ S. Michaud, S. Faucher, J.-S. Lauzon, M. Labb\'{e}, F. Grondin, F. Ferland, D. L\'{e}tourneau, F. Michaud are with the Department of Electrical Engineering and Computer Engineering, Interdisciplinary Institute for Technological Innovation (3IT), 3000 boul. de l'Universit\'{e}, Universit\'{e} de Sherbrooke, Qu\'{e}bec (Canada) J1K 0A5, {\texttt{\{Simon.Michaud,Samuel.Faucher2,Francois.Grondin2,\newline
Jean-Samuel.Lauzon,Mathieu.M.Labbe,\newline
Dominic.Letourneau,Francois.Ferland,\newline
Francois.Michaud\}@USherbrooke.ca}}}%
}

\newcommand{\smcomment}[1]{}
\newcommand{\sfcomment}[1]{}
\newcommand{\jslcomment}[1]{}
\newcommand{\mlcomment}[1]{}
\newcommand{\fgcomment}[1]{}
\newcommand{\ffcomment}[1]{}
\newcommand{\dlcomment}[1]{}
\newcommand{\fmcomment}[1]{}

\begin{document}

\maketitle
\thispagestyle{empty}
\pagestyle{empty}

\begin{abstract}
A microphone array can provide a mobile robot with the capability of localizing, tracking and separating distant sound sources in 2D, i.e., estimating their relative elevation and azimuth.
To combine acoustic data with visual information in real world settings, spatial correlation must be established.
The approach explored in this paper consists of having two robots, each equipped with a microphone array, localizing themselves in a shared reference map using SLAM. 
Based on their locations, data from the microphone arrays are used to triangulate in 3D the location of a sound source in relation to the same map.
This strategy results in a novel cooperative sound mapping approach using mobile microphone arrays. 
Trials are conducted using two mobile robots localizing a static or a moving
sound source to examine in which conditions this is possible.
Results suggest that errors under 0.3 m are observed when the relative angle between the two robots are above 30$\bm{^{\circ}}$ for a static sound source, while errors under 0.3 m for angles between 40$\bm{^{\circ}}$ and 140$\bm{^{\circ}}$ are observed with a moving sound source.
\end{abstract}


\section{INTRODUCTION}
\label{sec:introduction}

Over the last 20 years, there has been a growing interest in developing real-time on-board artificial audition capabilities on robots, with libraries like FlowDesigner \cite{letourneau2005flowdesigner}, HARK \cite{nakadai2010design}, ManyEars \cite{grondin2013manyears} and ODAS \cite{grondin2019lightweight}. 
In the recent five years, products equipped with microphone arrays (MAs) (e.g., Amazon Echo, Apple HomePod, Google Home
) opened a booming market, and development kits are now commonly available (e.g., ReSpeaker 6-MA, XMOS xCORE 7-MA, 8SoundsUSB 8-MA \cite{grondin2013manyears} and 16SoundsUSB 16-MA).  
Artificial audition technology aims at providing more natural interaction with connected devices such as mobile robots. 
Artificial audition on a mobile robot can enrich visual perception of the environment by helping to discover interesting elements in real world settings.
For instance, a person talking or an object making a sound can be used to draw the robot’s attention to something worth looking at more closely, associating the perceived sound with the image at that same location.
This information can then be used to make interesting multimodal associations \cite{sinapov-schenck-stoytchev2014multimodal}: face recognition can be used to identify the voice signature of a person, an image tagged to a ring may be designated as a telephone, etc. 

Doing so requires associating visual data and audio events in relation to the same reference frame. 
Visual SLAM (Simultaneous Localization and Mapping) can be used to generate a map of the environment to provide such reference frame of robots equipped with a MA doing sound source localization (SSL).
However, assuming that the sound sources are far from the robots compared to the MA aperture (a condition known as the far field effect), SSL only provides elevation and azimuth of sound sources \cite{sekiguchi2016online}.
Triangulating data from two or more MAs can be used to evaluate the 3D location of a sound source, as demonstrated in \cite{lauzon2017localization} when the locations of static MAs are known.
Using mobile MAs would make it possible, considering that the MA's positions are derived using SLAM and a shared reference map, to evaluate the 3D location of a sound source without having MAs placed in fixed positions.

    \begin{figure}[ht]
        \centering
        \includegraphics[width=\columnwidth]{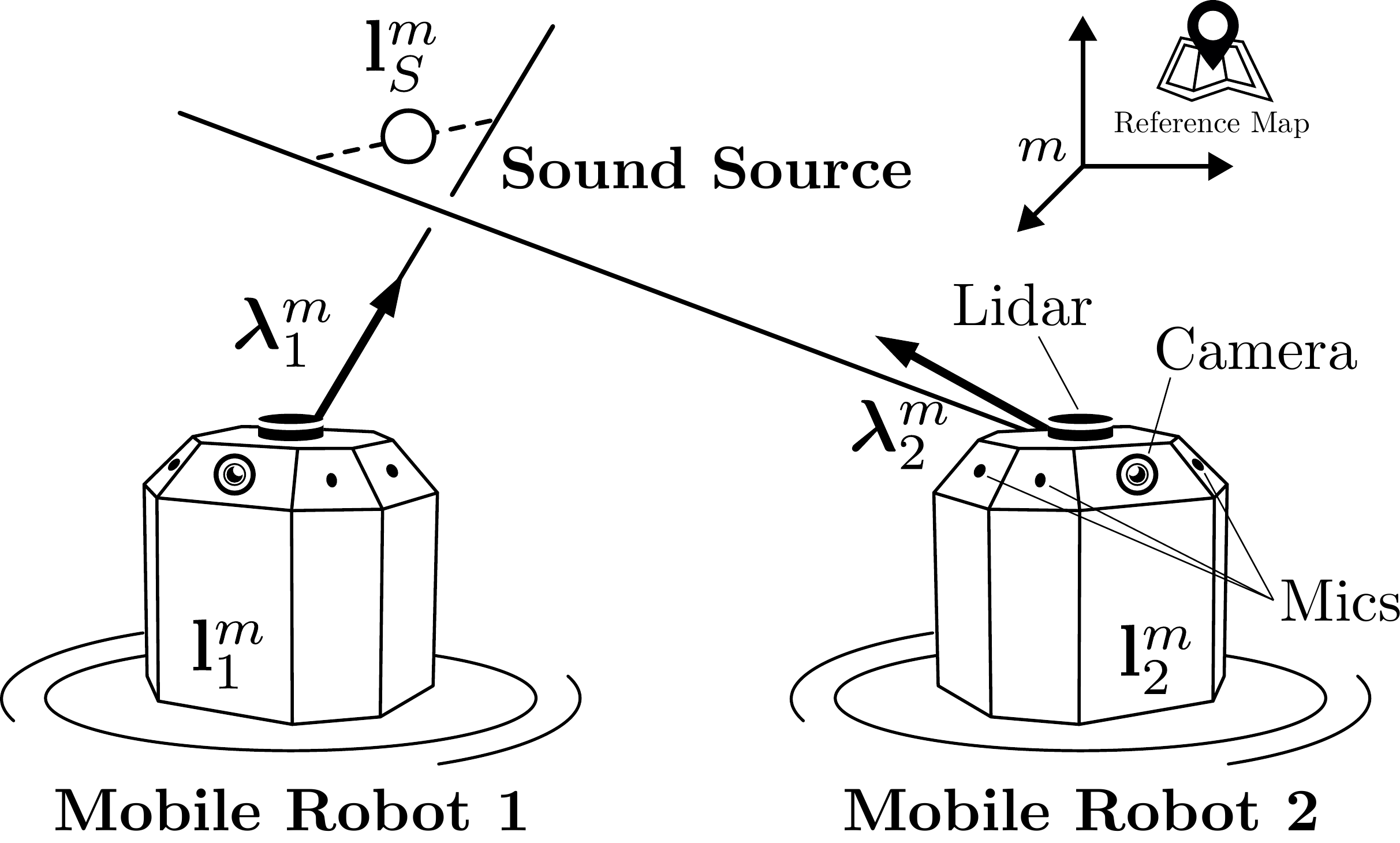}
        \caption{Localization of a sound source using mobile MAs}
        \label{fig:problem_statement}
    \end{figure}

Shown by Fig. \ref{fig:problem_statement}, this paper presents an approach addressing this research question using two mobile robots and one sound source.  
Each mobile robot is equipped with a lidar, a RGB-D camera and a MA. 
RTAB-Map (Real-Time Appearance-Based Mapping) \cite{labbe2018rtabmap}, a visual and lidar SLAM library, is used by the robots to localize themselves in a reference map $m$ of the environment. 
ODAS (Open embeddeD Audition System) \cite{grondin2019lightweight}, a sound source localization, tracking and separation library, is used to provide unit vectors $\bm{\lambda}_1 \in \mathcal{S}^2$ and $\bm{\lambda}_2 \in \mathcal{S}^2$ pointing in the direction of the sound source for each robot, where $\mathcal{S}^2 = \{v \in \mathbb{R}^3: \lVert v \rVert_2 = 1\}$, and $\lVert \dots \rVert_2$ stands for the $l_2$-norm. 
These open source libraries were chosen for convenience and to facilitate reproducibility of the results.
The closest intersection point of $\bm{\lambda}_1$ and $\bm{\lambda}_2$ is used to estimate the 3D location $\mathbf{l}_{S}^m$ of the sound source. 
The objective is identify the minimal conditions for 3D triangulation using mobile MAs is possible.
The paper is organized as follows. 
Section \ref{sec:relatedwork} provides an overview of work to situate our approach in relation to combining SSL and SLAM.
Section \ref{sec:system} presents the approach implemented.
Section \ref{sec:experiment} describes the experimental setup, followed by Section \ref{sec:results} with the observed results. 

\section{RELATED WORK}
\label{sec:relatedwork}

Rao-Blackwellized particle filter with Kalman filtering are commonly used for tracking sound sources.
For instance, Lin et al. \cite{lin2005cooperative} estimate the relative poses of a team of mobile robots, each robot equipped with a pair of microphones and emitting a specially-designed sound to simultaneously provide robot identification and the relative distances and bearing angles in 2D.
This acoustic data is combined with odometry and filtering is used to resolve the heading angle and the back-front ambiguities, implementing what is referred to as \textbf{cooperative acoustic robot localization} \cite{drioli2019acoustic}.
Teams of micro air vehicles (MAVs) equipped with 4-MAs use a similar concept with Extended Kalman Filtering (EKF) to position themselves in relation to a beacon MAV circling around a reference point in space while emitting continuous predefined acoustic chirps \cite{basiri2014mav}.

We identify three categories of approaches combining SSL with SLAM.
\textbf{Acoustic SLAM (aSLAM)} makes it possible to localize the trajectory of a MA on a mobile robot whilst estimating the acoustic map of surrounding sound sources \cite{evers2018aSLAM,evers2017tracking,evers2016localization}.
aSLAM basically exploits the movement of a MA to constructively triangulate over time the 3D cartesian location of sound sources from bearing-only 2D Direction-of-Arrival (DoA) measurements, estimating the robot trajectory from the apparent displacement of sound sources observed from multiple positions.
aSLAM performance is therefore affected by the trajectory followed by the MA in relation to the sound sources.
Only validated in simulation, this approach is limited to a single robot with mapping referenced to the robot's initial position, and requires at least two sound sources to work. 
Similar limitations apply to \cite{schymura2018potential,nguyen-colas-charpillet2016} which adopt a similar approach using Kalman filtering.
In the same category, Sasaki et al. \cite{sasaki2010iros,sasaki2006iros} derive 2D positions of multiple sound sources using a 32-MA and sound observations over the last 2 sec, and sound source categorization is used to remove undesirable cross points.
In more recent work, Sasaki et al. \cite{sasaki2016iros} designed a hand-held unit equipped with a 3D lidar and Inertial Measurement Unit (IMU) for SLAM, and using HARK with a 32-MA for SSL.
Particle filtering from data taken over time (from 7.5 to 15 sec) with the hand-held unit moving (rotation, displacement) provides 3D positions of two sound sources. 

An extension to aSLAM is \textbf{audio-visual SLAM (AV-SLAM)}, exploiting acoustic and visual features \cite{chau2019audio} for human tracking.
In \cite{chau2019audio}, validation is done using one robot equipped with a 8-MA running HARK and a RGB camera moving in a straight line over 2.5 m in front of a stationary human sound source, providing only 2D localization. 
Bayram and Ince \cite{bayram2015sii} present audio-visual multi-person 2D tracking by doing sensor fusion of a SSL module with a visual face recognition module.
Results are presented using two Kinect cameras and a 7-MA running HARK.

Finally, the concept of \textbf{audio-based SLAM} \cite{sekiguchi2016online} involves considering the SSL problem as a SLAM problem.
For instance, the FastSLAM \cite{thrun2004fastslam} algorithm is used to estimate the time offset and position of robots equipped with MAs and the position of sound sources \cite{hu2009icra}.
Sekiguchi et al. \cite{sekiguchi2016online} use FastSLAM with static MA-equipped robots to consider the MAs as one big array.
Results using HARK and three static robots with 8-MAs in an anechoic chamber and two moving talkers are provided.
Audio-based SLAM has also been used for online calibration of asynchronous MAs \cite{miura2011calibration,su-vidal-calleja-miro2015-7354165} and for optimizing the relative positions of multiple mobile robots with MAs for cooperative sound source separation \cite{sekiguchi2015cooperative}.

\section{Sound Source Mapping Using Mobile MAs}
\label{sec:system}

\begin{figure*}[ht]
      \vspace{0.2cm}
      \centering
      \includegraphics[width=\textwidth]{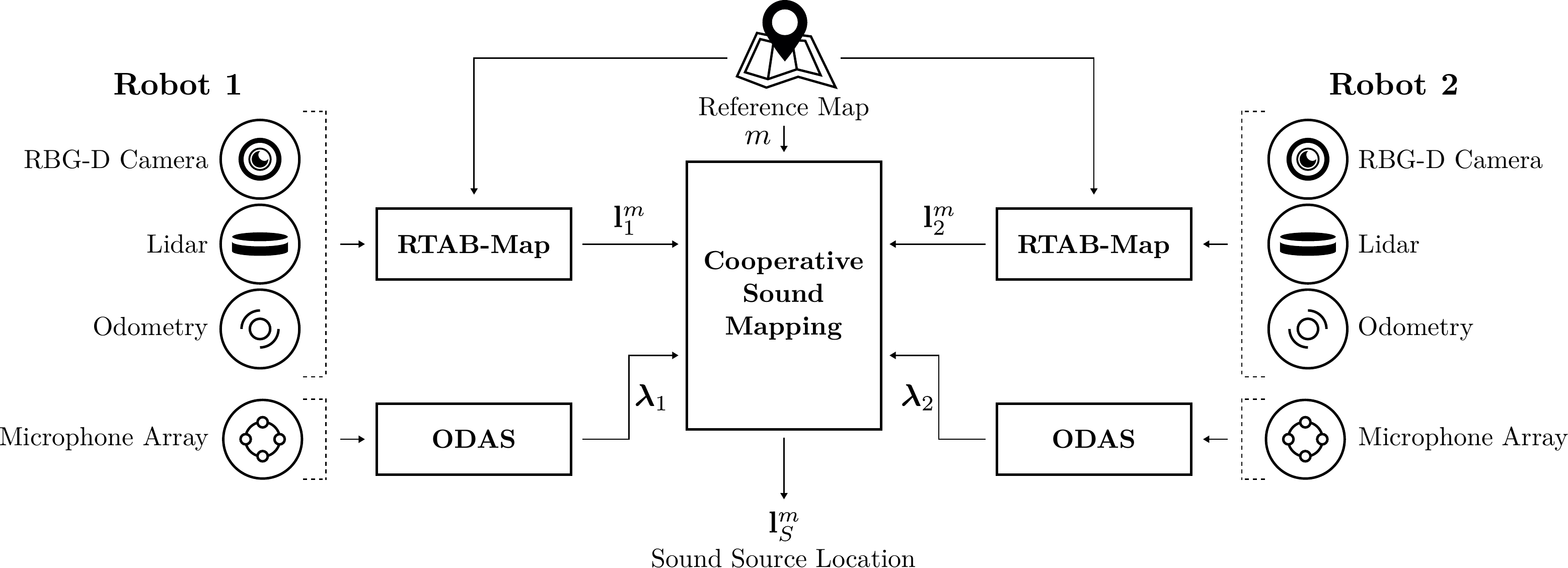}
      \caption{Architecture diagram of the cooperative sound mapping approach}
      \label{fig:system_overview}
   \end{figure*}

Our approach aims at using two mobile MAs to provide instantaneous 3D location of a sound source.
In relation to Section \ref{sec:relatedwork}, our approach differs by having mobile MAs, localized using SLAM based on $m$, triangulate sound source locations in 3D also in relation to $m$. 
The concept can be designated as \textbf{cooperative sound mapping}, illustrated by the architecture diagram presented by Fig. \ref{fig:system_overview}. 
Each mobile robot $i$ is equipped with a lidar, a RGB-D camera and onboard odometry to localize its location $\mathbf{l}^m_i \in \mathbb{R}^7$ (3D position and a quaternion for rotation) in relation to a reference map $m$ using RTAB-Map. 
Each robot $i$ is also equipped with a 16-MA and uses ODAS to do sound source localization, tracking and separation.
ODAS provides a 3D unit vector $\bm{\lambda}_i \in \mathcal{S}^2$ pointing in the direction of the sound source with respect to the robot.
Using data from two mobile robots ($i = \{1,2\}$), the Cooperative Sound Mapping module triangulates the position of the sound source in relation to the reference map.

\subsection{RTAB-Map}
\label{sec:rtab-map}

RTAB-Map (Real-Time Appearance-Based Mapping) 
\cite{labbe-michaud2017} is an open source library\footnote{http://introlab.github.io/rtabmap/} implementing graph-based SPLAM (Simultaneous Planning, Localization And Mapping) \cite{stachniss2009robotic} i.e., the ability to simultaneously map an environment, localize itself in it and plan paths.
RTAB-Map provides the robots' positions and orientations, denoted as $\mathbf{l}_{1}^m$ and $\mathbf{l}_{2}^m$, respectively.
RTAB-Map uses a combination of odometry, lidar and camera to robustly create a map and to localize in it. The lidar is used to create the 2D occupancy grid map for obstacle avoidance and path planning. Appearance-based loop closure detection and localization are done with visual features extracted from the RGB image of the RGB-D camera using a bag-of-words approach. By estimating the 3D positions of visual features using the depth image, a position and orientation against the map can be computed. The localization is then refined using the lidar to improve accuracy when environments are lacking visual features but has a lot of geometry (which is often the case indoor).

\subsection{ODAS}
\label{sec:odas}

ODAS \cite{grondin2019lightweight} is an open source library\footnote{http://odas.io} performing sound source localization, tracking and separation. 
ODAS generates the DoA for each of the two robots ($i=\{1,2\}$), denoted as $\bm{\lambda}_1$ and $\bm{\lambda}_2$. 
This library relies on a localization method called Steered Response Power with Phase Transform based on Hierarchical Search with Directivity model and Automatic calibration (SRP-PHAT-HSDA).
The proposed approach decomposes the search space in coarse and fine grids, which speeds up the search in 3D for the DoA of one or many sound sources.
Localization generates noisy potential sources, which are then filtered with a tracking method based on a modified 3D Kalman filter (M3K) that generates one or many tracked sources.
Sound sources are then filtered and separated using directive geometric source separation (DGSS) to focus the robot's attention only on the target sound source, and ignore ambient noise. 
This library also models microphones as sensors with a directive polar pattern, which improves sound source localization, tracking and separation when the direct path between microphones and the sound sources is obstructed by the robot's body.

In this work, ODAS is configured to return the loudest sound source DoA per robot, denoted as $\bm{\lambda}_1$ and $\bm{\lambda}_2$ for robots $i = \{1,2\}$.
Time synchronization between $\bm{\lambda}_1$ and $\bm{\lambda}_2$ is facilitated using ODAS' tracking module output because the DoAs are smoothed over time.

\subsection{Cooperative Sound Mapping}
\label{sec:CSM}


The Cooperative Sound Mapping module combines the information provided by the two robots to determine the location of the sound source $\mathbf{l}_{S}^m \in \mathbb{R}^3$.
It first rotates the DoA for each robot in relation to its orientation to derive the vectors $\bm{\lambda}_1$ and $\bm{\lambda}_2$.
In 3D space, $\bm{\lambda}_1$ and $\bm{\lambda}_2$ rarely intersect each other.
The estimation of $l_S^m$ is derived by finding the smallest distance between $\bm{\lambda}_1$ and $\bm{\lambda}_2$, as represented by the dotted line in Fig. \ref{fig:problem_statement}.
Using the Ray to Ray algorithm \cite{schneider2002geometric} as in \cite{lauzon2017localization}, the sound source position is estimated using (\ref{eq:ray_to_ray}):

\begin{equation}
    \mathbf{l}^m_S = \frac{1}{2}(\mathbf{l}^m_1 + G_{1}\bm{\lambda}_1 + \mathbf{l}^m_2 + G_{2}\bm{\lambda}_2)
    \label{eq:ray_to_ray}
\end{equation}
where the expressions $G_1$ and $G_2$ are given as follows:
\begin{equation}
    G_1 = \frac{(\bm{\lambda}_1 \cdot \bm{\lambda}_2)(\bm{\lambda}_2 \cdot (\mathbf{l}_1^m - \mathbf{l}_2^m))-\bm{\lambda}_1 \cdot (\mathbf{l}_1^m - \mathbf{l}_2^m)}{1-(\bm{\lambda}_1 \cdot \bm{\lambda}_2)^2}
\end{equation}
\begin{equation}
    G_2 = \frac{(\bm{\lambda}_1 \cdot \bm{\lambda}_2)(\bm{\lambda}_1 \cdot (\mathbf{l}_2^m - \mathbf{l}_1^m)) - \bm{\lambda}_2 \cdot (\mathbf{l}_2^m - \mathbf{l}_1^m)}{1-(\bm{\lambda}_1 \cdot \bm{\lambda}_2)^2}
\end{equation}
and $(\cdot)$ stands for the dot product.

\section{Experimental Methodology}
\label{sec:experiment}

\begin{figure}[ht!]
    \centering
    \includegraphics[width=\linewidth]{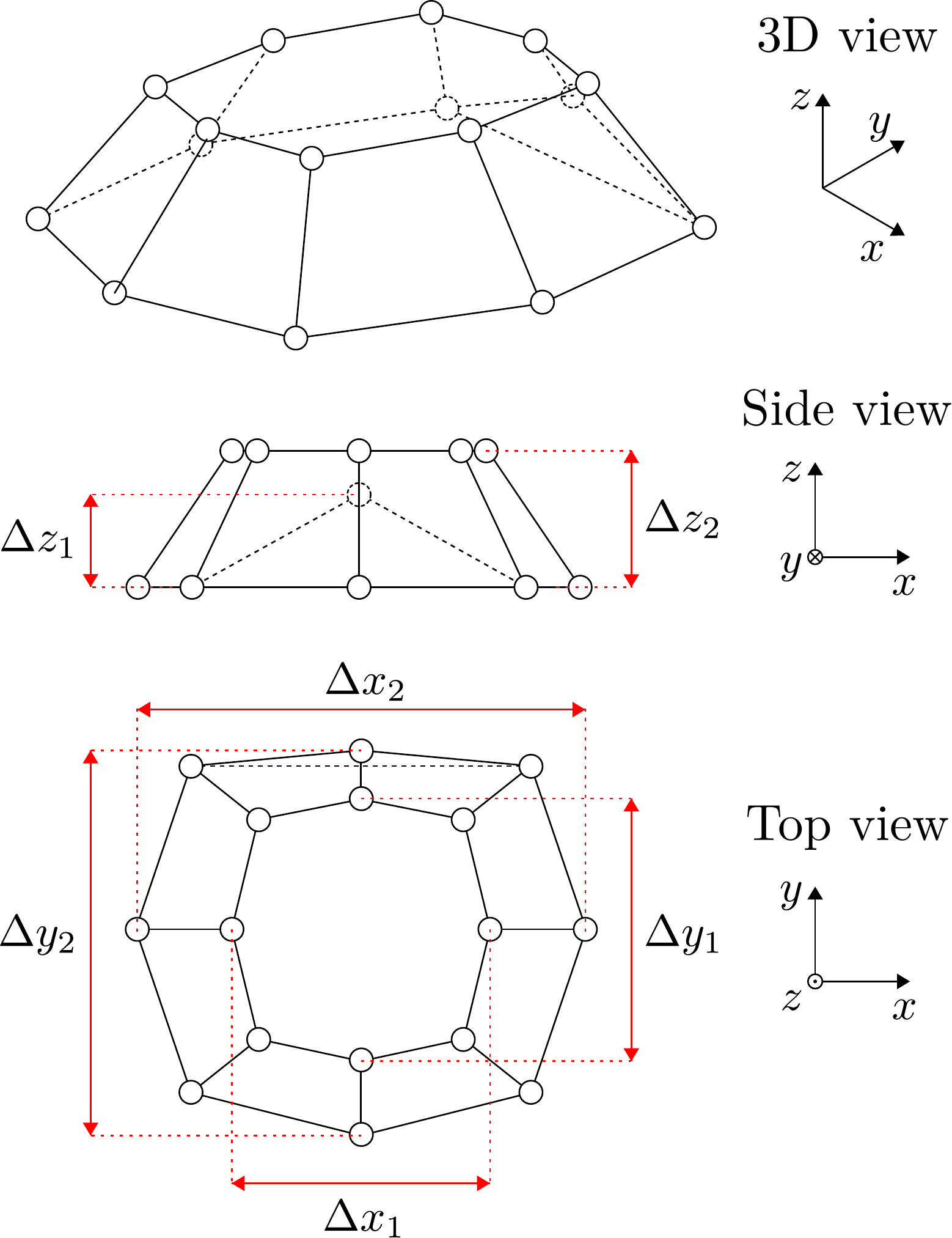}
    \caption{MA configurations}
      \label{fig:micarray}
\end{figure}

Two SecurBot\footnote{https://github.com/introlab/securbot} mobile robots are used: one equipped with a Jetson Nano core mounted on a TurtleBot2 base, and the other with a Jetson TX2 installed on a modified Pioneer2-DX base. 
Both robots are equipped with a 16SoundsUSB\footnote{https://github.com/introlab/16SoundsUSB} MA, an Intel Realsense D435 camera and a RP Lidar. 
Each MA is located 0.48 m above the ground and provides synchronous acquisition of microphone signals through USB to the robot's computer. 
Figure \ref{fig:micarray} and Table \ref{tab:micposition} present the MA configuration of each robot.

\begin{table}[]
    \renewcommand*{\arraystretch}{1.5}
    \centering
    \caption{Position of the microphones on the MA (cm)}
    \begin{tabular}{|c||cccccc|}
        \hline
        Dimensions & $\Delta x_1$ & $\Delta x_2$ & $\Delta y_1$ & $\Delta y_2$ & $\Delta z_1$ & $\Delta z_2$ \\
        \hline
        Pioneer2-DX & $19.1$ & $33.2$ & $24.5$ & $36.0$ & $2.5$ & $4.0$ \\
        TurtleBot2 & $18.5$ & $31.0$ & $19.3$ & $28.4$ & $2.6$ & $3.9$ \\
        \hline
    \end{tabular}
    \label{tab:micposition}
\end{table}

The experiments are conducted in a 150 m{$^2$} room filled with different objects to provide visual features for SLAM. 
The reference map is created using RTAB-Map with one of the robot. 
The room has a reverberation level of $RT_{60} = 600$ msec and no background noise.
ODAS is configured with similar parameters as those in \cite{grondin2019lightweight}, except for the covariance component of the observation noise matrix in the Kalman filter of ODAS' tracking module, which is increased to $\sigma_R^2 = 0.01$ for more sensitivity to sound source acceleration, which influences $\lambda$.

\begin{figure}[!ht]
\centering
\vspace{2mm}
\begin{subfigure}{\columnwidth}
\includegraphics[width=\linewidth]{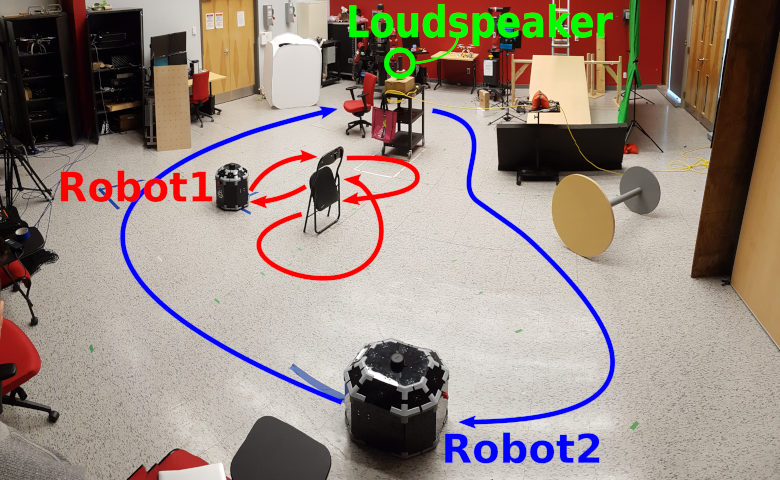}
\caption{Static sound source}
\label{fig:setup_robot_oma}  
\vspace{5pt}
\end{subfigure}
\begin{subfigure}{\columnwidth}
\includegraphics[width=\columnwidth]{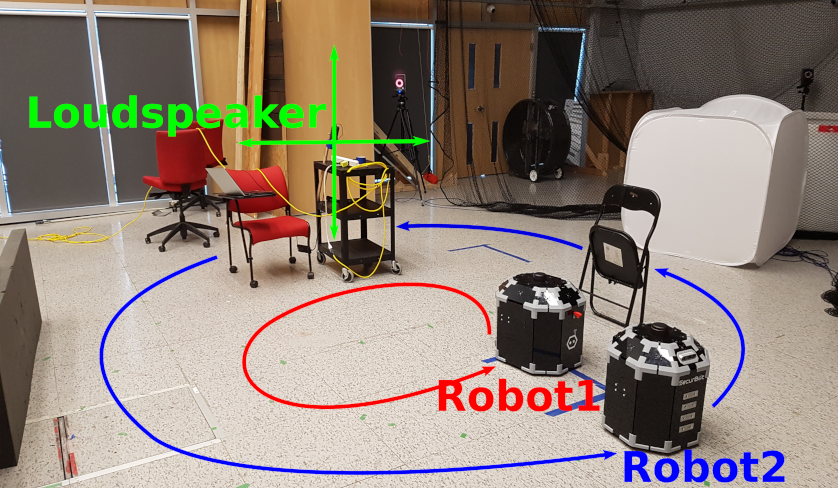}
\caption{Moving sound source}
\label{fig:setup_robot_cma}
\end{subfigure}
\caption{Experimental conditions with the two SecurBot robots}
\label{fig:setup_robot}
\end{figure}

Trials are conducted with the sound source located at a static location at 1.12 m of height 
or by being manually moved horizontally and vertically, with its location being monitored using a Vicon motion system.  
The sound source is a loudspeaker generating white noise, with perceived amplitude ranging from 25 to 30 dB over 3 m.
Robots move around the sound source by following preset trajectories defined in relation to the reference map. 
Figure \ref{fig:setup_robot} shows the trajectories followed by the robots using ROS's navigation stack \cite{quigley2009ros}\footnote{http://www.ros.org}. 
These trajectories have the robots move from 0 to 2 m/sec and are set to avoid collisions between the two robots and to cover a variety of DoA configurations in relation to the sound source.
A remote laptop computer, also running ROS, monitors the trials and records the localization and audio data. 
RViz is used to display the position of the robots, the map, the DoAs, the estimated and the known locations of the sound source. 
RViz also displays Root Square Error (RSE) in meter between $l_S^m$ and the actual sound source location with colored dots ranging from green to black from 0 to 0.5 m, and to red as it increases.

\section{Results}
\label{sec:results}

\begin{figure}[!ht]
    \centering
    \includegraphics[width=0.9\columnwidth]{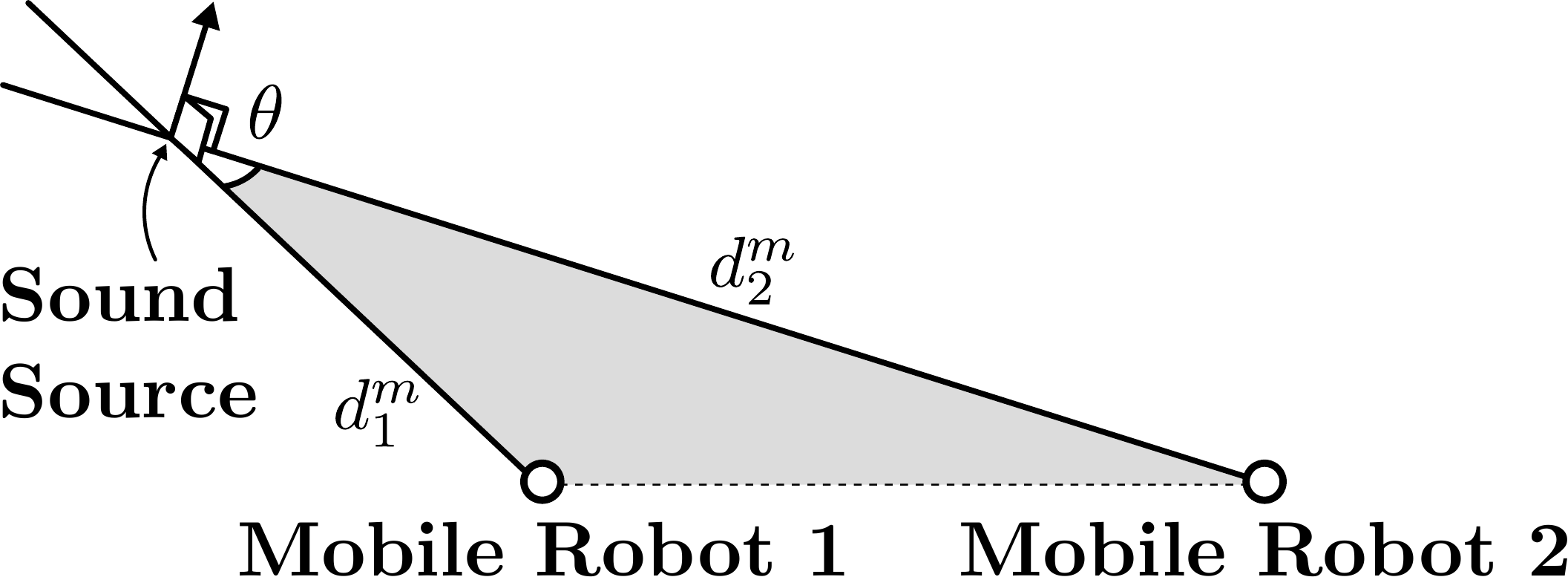}
    \vspace{5pt}
    \caption{Definition of $d_1^m$, $d_2^m$ and $\theta$}
    \label{fig:results_triangle}
\end{figure}

RSE is examined in relation to $d_1^m$ and $d_2^m$, the distances between the robots and the sound source, and the angle $\theta$ as defined by Fig. \ref{fig:results_triangle}.
$\theta$ is the angle between the lines from the theoretical location of the sound source to $l_1^m$ and $l_2^m$ in 3D, as shown by Fig. \ref{fig:problem_statement}.  

\begin{figure*}
 \vspace{1mm}
   \centering
    \begin{subfigure}{0.32\textwidth}
        \centering
        \resizebox{0.97\linewidth}{!}{%
            \begin{tikzpicture}
                \begin{axis}[axis y line*=left, xmin=0, xmax=112, ymin=0, ymax=5, xlabel=Time (sec), ylabel=$d_1^m$ (m) \textcolor{blue}{---} $d_2^m$ (m) \textcolor{green}{---}]
                    \addplot[mark=none, blue] table [x=t, y=d1, col sep=comma] {data_static_dist.csv};
                    \addplot[mark=none, green] table [x=t, y=d2, col sep=comma] {data_static_dist.csv};
                \end{axis}
                \begin{axis}[axis x line=none, axis y line*=right, xmin=0, xmax=112, ymin=0, ymax=10, xlabel=Time (sec), ylabel=RSE (m) \textcolor{red}{---}]
                    \addplot[mark=none, red] table [x=t, y=rse, col sep=comma] {data_static_dist.csv};
                \end{axis}
            \end{tikzpicture}
        }
        \caption{$d_1^m$, $d_2^m$ and RSE over time}\label{fig:results_static_dis}
    \end{subfigure}
    \begin{subfigure}{0.32\textwidth}
        \centering
        \resizebox{\linewidth}{!}{%
            \begin{tikzpicture}
                \begin{axis}[axis y line*=left, xmin=0, xmax=112, ymin=0, ymax=180, xlabel=Time (sec), ylabel=$\theta$ ($^{\circ}$) \textcolor{blue}{---}]
                    \addplot[mark=none, blue] table [x=t, y=theta, col sep=comma] {data_static_time.csv};
                \end{axis}
                \begin{axis}[axis x line=none, axis y line*=right, xmin=0, xmax=112, ymin=0, ymax=10, ylabel=RSE (m) \textcolor{red}{---}]
                    \addplot[mark=none, red] table [x=t, y=rse, col sep=comma] {data_static_time.csv};
                 \end{axis}    
            \end{tikzpicture}
        }
        \caption{$\theta$ and RSE over time}\label{fig:results_static_angle}
    \end{subfigure}
    \begin{subfigure}{0.32\textwidth}
        \centering
        \resizebox{0.85\linewidth}{!}{%
            \centering
            \begin{tikzpicture}
                \begin{axis}[xmin=0, xmax=180, ymin=0, ymax=1, xlabel=$\theta$, ylabel=RSE (m) \textcolor{black}{---}]
                    \addplot+[color=black,mark options={fill=black}, error bars/.cd,y fixed, y dir=both, y explicit] table [x=theta, y=rse_mean, y error=rse_std, col sep=comma] {data_static_stats.csv};
                \end{axis}
            \end{tikzpicture}
        }
        \caption{RSE in relation to $\theta$}\label{fig:results_static_rse_angle}
    \end{subfigure}
    \caption{Trial with a static sound source}
    \label{fig:results_static}
\end{figure*}
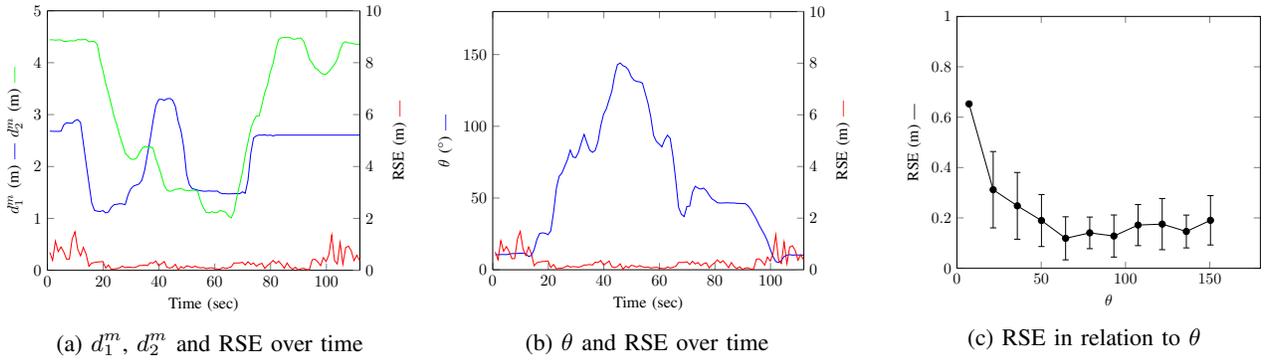

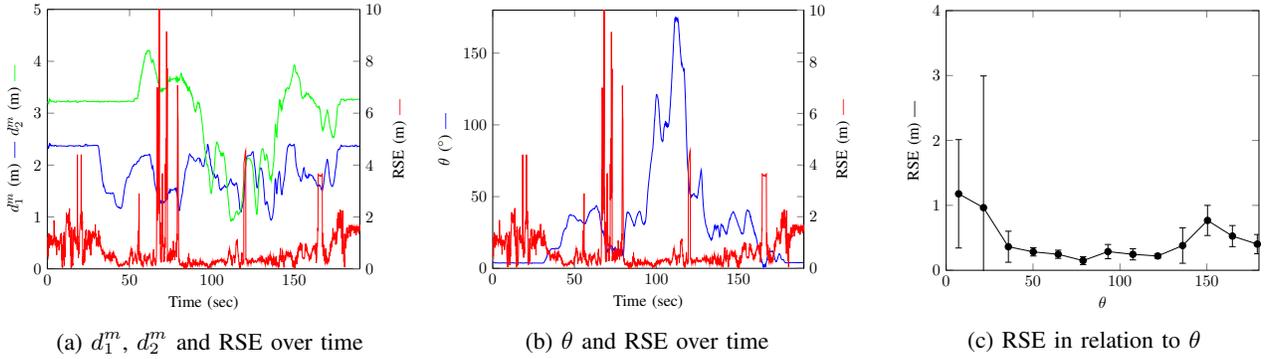
\begin{figure*}
    \centering
    \begin{subfigure}{0.32\textwidth}
        \centering
        \resizebox{0.97\linewidth}{!}{%
            \begin{tikzpicture}
                \begin{axis}[axis y line*=left, xmin=0, xmax=190, ymin=0, ymax=5, xlabel=Time (sec), ylabel=$d_1^m$ (m) \textcolor{blue}{---} $d_2^m$ (m)  \textcolor{green}{---}]
                    \addplot[mark=none, blue] table [x=t, y=d1, col sep=comma] {data_moving_dist.csv};
                    \addplot[mark=none, green] table [x=t, y=d2, col sep=comma] {data_moving_dist.csv};
                \end{axis}
                \begin{axis}[axis x line=none, axis y line*=right, xmin=0, xmax=190, ymin=0, ymax=10, xlabel=Time (sec), ylabel=RSE (m) \textcolor{red}{---}]
                    \addplot[mark=none, red] table [x=t, y=rse, col sep=comma] {data_moving_dist.csv};
                \end{axis}
            \end{tikzpicture}
        }
        \caption{$d_1^m$, $d_2^m$ and RSE over time}\label{fig:results_moving_dis}
    \end{subfigure}
    \begin{subfigure}{0.32\textwidth}
        \centering
        \resizebox{\linewidth}{!}{%
            \begin{tikzpicture}
                \begin{axis}[xmin=0, xmax=190, ymin=0, ymax=180, xlabel=Time (sec), ylabel=$\theta$ ($^{\circ}$) \textcolor{blue}{---}]
                    \addplot[mark=none, blue] table [x=t, y=theta, col sep=comma] {data_moving_time.csv};
                \end{axis}
                \begin{axis}[axis x line=none, axis y line*=right, xmin=0, xmax=190, ymin=0, ymax=10, ylabel=RSE (m) \textcolor{red}{---}]
                    \addplot[mark=none, red] table [x=t, y=rse, col sep=comma] {data_moving_time.csv};
                 \end{axis}    
            \end{tikzpicture}
        }
        \caption{$\theta$ and RSE over time}\label{fig:results_moving_angle}
    \end{subfigure}
    \begin{subfigure}{0.32\textwidth}
        \centering
        \resizebox{0.85\linewidth}{!}{%
            \begin{tikzpicture}
                \begin{axis}[xmin=0, xmax=180, ymin=0, ymax=4, xlabel=$\theta$, ylabel=RSE (m) \textcolor{black}{---}]
                    \addplot+[color=black,mark options={fill=black}, error bars/.cd,y fixed, y dir=both, y explicit] table [x=theta, y=rse_mean, y error=rse_std, col sep=comma] {data_moving_stats.csv};
                \end{axis}
            \end{tikzpicture}
        }
        \caption{RSE in relation to $\theta$}\label{fig:results_moving_rse_angle}
    \end{subfigure}
    \caption{Trial with a moving sound source}
    \label{fig:results_moving}
\end{figure*}

\begin{figure}[!ht]
\includegraphics[width=\columnwidth]{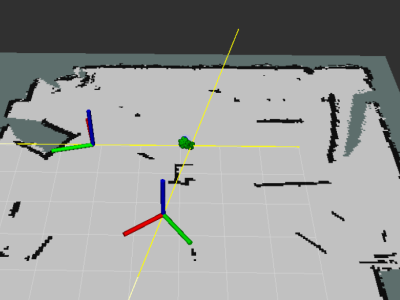}
\caption{Illustration in RViz of a case with $\theta > 15^{\circ}$}
\label{fig:setup_robot_snapshot_cross}
\end{figure}

\begin{figure}[!ht]
\vspace{2mm}
\includegraphics[width=\linewidth]{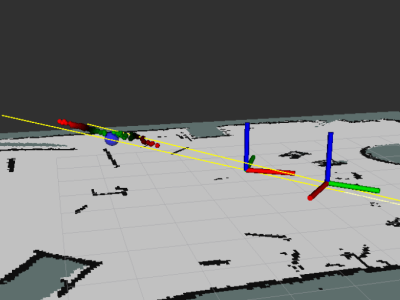}
\caption{Illustration in RViz of a case with $\theta$ small}
\label{fig:setup_robot_snapshot_para}  
\end{figure}

Figure \ref{fig:results_static} and Fig. \ref{fig:results_moving} summarizes the observations made during a trial with a static sound source and a moving sound source. 
For the static sound source, data is recorded at 10 Hz.
Fig. \ref{fig:results_static_dis} to Fig. \ref{fig:results_static_rse_angle} illustrate that RSE remains lower than approximately 0.3 m with negligible variance for $\theta$ higher than approximately $30^{\circ}$, regardless of the distances between the robots and the sound source, as illustrated by RViz in Fig. \ref{fig:setup_robot_snapshot_cross}. 
These results are confirmed by Fig. \ref{fig:results_static_rse_angle} which represents, over intervals of 15$^{\circ}$, the average RSE and its standard deviation.  
For smaller $\theta$ (which occur at the first and the last $\sim$18 sec of the static sound source trial), $\bm{\lambda}_1$ and $\bm{\lambda}_2$ are becoming parallel, and small changes in $\theta$ lead in larger errors.
When parallel, the denominator of (\ref{eq:ray_to_ray}) results in a division by zero. 
So when $\theta$ is small, the closest intersection point found between $\bm{\lambda}_1$ and $\bm{\lambda}_2$ quickly changes as the denominator of (\ref{eq:ray_to_ray}) comes closer to zero.
Figure \ref{fig:setup_robot_snapshot_para} illustrates this situation.
One alternative to limit such occurrences would be to use a third robot, to actively reposition the robots to have $\theta$ higher than $15^{\circ}$, or to use robots with MAs at different heights.

The static sound source trial limits the possible range for $\theta$ because the sound source is located higher than MAs.
The moving sound source trial makes it possible to have $\theta$ change from 0 to 180$^{\circ}$.
Fig. \ref{fig:results_moving_dis} to Fig. \ref{fig:results_moving_angle} present data (recorded at 100 Hz to synchronize with the Vicon system).
Between 0 to 30 sec, the robots are immobile and near each other, and the sound source is moving.
Because $\theta$ is small (9$^{\circ}$), large errors are observed as explained with the static condition.
From 30 to 61 sec, robots are moving away from each other, with one coming closer to the sound source being static: as $\theta$ increases, RSE decreases. 
The peak at 55 sec is caused by $\sigma_R^2$, overshooting $l_S^m$ based on the influence of the sound source acceleration as robots are near each other and the sound source starts moving.
From 61 to 80 sec, large peaks occur as $\theta$ is small (i.e., $\bm{\lambda}_1$ and $\bm{\lambda}_2$ are almost parallel) and the robots' motion makes $\theta$ change quickly. 
From 80 to 155 sec, $\theta$ is sufficient to have small RSE except when $\theta$ is near 180$^{\circ}$ (111 sec) and at 120 sec when there was an obstacle between one robot and the sound source. 
For the remaining time of the trial, $\theta$ rapidly decreases toward 0, creating RSE peaks and larger errors. 
Figure \ref{fig:results_moving_rse_angle} suggest that for $\theta$ between $\sim$40$^{\circ}$ to $\sim$140$^{\circ}$, the RSE is lower than 0.3 m.

\section{CONCLUSION}
\label{sec:conclusion}
This paper validates the concept of cooperative sound mapping by demonstrating, using RTAB-Map ad ODAS, that it is possible to derive the 3D location of a sound source using mobile MAs. 
Results show that the capability of approximating the location of the sound source from the closest intersection point found between $\bm{\lambda}_1$ and $\bm{\lambda}_2$ is influenced by $\theta$ and the sensitivity of sound source tracking.
This could be filtered out using a Kalman filter, as we observed in our preliminary trials. 
As the experiments presented in the paper involve the ideal case of only having one constant sound source, the next steps in our work involve extending the approach to use two and more MAs for online simultaneous localization of multiple intermittent sound sources in noisy and reverberation conditions, and coordinate the positioning of the mobile robots to provide reliable 3D location measurements according to their positions in relation to the sound sources.
We believe that directly using the output of sound source tracking instead of SSL will simplify the overall complexity for cooperative sound mapping, targeting onboard centralized or distributed processing. 



\bibliographystyle{IEEEtran}
\bibliography{references}

\end{document}